\documentclass[reprint,aps,prc,superscriptaddress,unsortedaddress,showpacs,groupedaddress,superscriptaddress,preprintnumbers]{revtex4-1}
\usepackage{graphicx}
\usepackage{dcolumn}
\usepackage{amsmath}
\usepackage{amssymb}
\usepackage{amsfonts}
\usepackage{times}
\usepackage{comment}
\usepackage{mathrsfs}
\usepackage{multirow}	
\usepackage{bm}
\usepackage{braket}
\usepackage[T1]{fontenc}
\usepackage[utf8]{inputenc}

\usepackage{color}
\usepackage{ulem}

\def\fun#1#2{\lower3.6pt\vbox{\baselineskip0pt\lineskip.9pt
\ialign{$\mathsurround=0pt#1\hfil##\hfil$\crcr#2\crcr\sim\crcr}}}

\begin{document}
\preprint{NITEP 24}
\title{
Toward a reliable description of $\bm{(p,pN)}$ reactions in the distorted-wave impulse approximation}

\author{Nguyen Tri Toan Phuc}
\email{nguyentritoanphuc@yahoo.com} 
\affiliation{Department of Nuclear Physics, Faculty of Physics and Engineering Physics, University of Science, VNU-HCM, 227 Nguyen Van Cu, District 5, 700000, Ho Chi Minh City, Vietnam}

\author{Kazuki Yoshida}
\email[]{yoshida.kazuki@jaea.go.jp}
\affiliation{Advanced Science Research Center, Japan Atomic Energy Agency, Tokai, Ibaraki 319-1195, Japan}

\author{Kazuyuki Ogata}
\email[]{kazuyuki@rcnp.osaka-u.ac.jp}
\affiliation{Research Center for Nuclear Physics (RCNP), Osaka
University, Ibaraki 567-0047, Japan}
\affiliation{Department of Physics, Osaka City University, Osaka 558-8585, Japan}
\affiliation{Nambu Yoichiro Institute of Theoretical and Experimental Physics (NITEP), Osaka City University, Osaka 558-8585, Japan}


\begin{abstract}

\begin{description}
\item[Background] Proton-induced nucleon knockout $(p,pN)$ reactions have been successfully used to study the single-particle nature of stable nuclei in normal kinematics with the distorted-wave impulse approximation (DWIA) framework. Recently, these reactions have been applied to rare-isotope beams at intermediate energies in inverse kinematics to study the quenching of spectroscopic factors.  

\item[Purpose] Our goal is to investigate the effects of various corrections and uncertainties within the standard DWIA formalism on the $(p,pN)$ cross sections. The consistency of the extracted reduction factors between DWIA and other methods is also evaluated.

\item[Method] We analyze the $(p,2p)$ and $(p,pn)$ reactions data measured at the R$^3$B/LAND setup at GSI for carbon, nitrogen, and oxygen isotopes in the incident energy range of 300--450 MeV/u. Cross sections and reduction factors are calculated by using the DWIA method. The transverse momentum distribution of the $^{12}$C($p$,$2p$)$^{11}$B reaction is also investigated.    

\item[Results] We have found that including the nonlocality corrections and the M\o ller factor affects the cross sections considerably. The proton-neutron asymmetry dependence of reduction factors extracted by the DWIA calculation is very weak and consistent with those given by other reaction methods and \textit{ab initio} structure calculations.   

\item[Conclusions] The results found in this work provide a detailed investigation of the DWIA method for $(p,pN)$ reactions at intermediate energies. They also suggest that some higher-order effects, which is essential for an accurate cross-section description at large recoil momentum, is missing in the current DWIA and other reaction models. 

\end{description}
\end{abstract}


\maketitle

\section{Introduction}
In the last fifty years, nucleon knockout reactions induced by intermediate energy protons of the type $(p,pN)$ in normal (forward) kinematics have been one of the most successful tools for studying the single-particle nature of stable nuclei \cite{Jac66,Jac73,Kit85,Wak17}. These reactions are sometimes referred to as quasifree scattering due to the dominance of scattering processes between the proton and the knocked-out nucleon. The most widely used theoretical approach to analyze these reactions is the distorted-wave impulse approximation (DWIA) method of Chant and Roos \cite{Cha77,Cha83}. A detailed review of this method including its uncertainties and applications to existing forward-kinematics $(p,pN)$ data has been reported in Ref.~\cite{Wak17}.         

With the availability of radioactive beams at energies up to 450 MeV/u, there has been a renewed interest to elucidate single-particle properties of unstable nuclei by using nucleon knockout reactions on the hydrogen target in inverse kinematics. The DWIA method has first been theoretically applied to inverse kinematics $(p,pN)$ reactions with the scattering wave functions treated with the eikonal approximation \cite{Aum13,Oga15} and later in (standard) partial-wave expansion form \cite{Yos18}. The latter method was then successfully applied in several experimental studies \cite{Oli17,Ele19,Tan19,Chen19,Kaw18} carried out at RIKEN, Japan. However, an extensive investigation of the sensitivity of calculated cross sections to various choices of inputs and corrections (as in Ref.~\cite{Wak17}) was not done in these studies.

Besides DWIA, other theoretical methods such as the three-body Faddeev equation in the Alt-Grassberger-Sandhas formulation (FAGS) \cite{Cre08,Cre19} and the transfer-to-the-continuum (TC) method \cite{Mor15} have also been used to analyze the $(p,pN)$ experimental data measured in inverse kinematics. Although there is a good consistency between DWIA and TC formalism for the specific $^{15}$C($p$,$pn$)$^{14}$C reaction \cite{Yos18}, a systematic comparison between the DWIA (eikonal and partial-wave form), TC, and FAGS analyses on the actual experimental data is essential to determine the range of applicability for $(p,pN)$ reactions.

Among the possible uses of proton-induced nucleon knockout reactions, the study of quenching single-particle strength and its proton-neutron asymmetry dependence is one of the most important subjects. A reduction of 30\%--40\% with respect to the independent-particle model (IPM) limit in spectroscopic factors (SFs) deduced from $(e,e'p)$ experiments was observed at NIKHEF \cite{Lap93}. This quenching of the SF, quantified as the reduction factor $R_s$, is due to the lack of short-range (including tensor) and long-range correlations in the IPM and standard shell model (SM) calculations \cite{Pan97,Dic04,Sick07,Bar09,Jen11,Cip15}. 

Systematic analysis of nucleon removal reactions on light composite targets ($^{9}$Be and $^{12}$C) \cite{Gade08,Tos14} suggests a strong dependence on the proton-neutron asymmetry defined as ${\Delta S=S_p-S_n}$ $(\Delta S=S_n-S_p)$ for proton (neutron) removal, where $S_p$ $(S_n)$ is the proton (neutron) separation energies. However, such strong dependence cannot be observed in systematic nucleon-transfer studies \cite{Lee06,Kay13,Fla13,Fla18,Xu19}. A weak dependence of $R_s$ on $\Delta S$ is also supported by \textit{ab initio} coupled-cluster (CC) \cite{Jen11} and self-consistent Green's function (SCGF) \cite{Cip15} calculations.             
 
Very recently, a series of $(p,pN)$ measurements for carbon-, nitrogen-, and oxygen-isotope beam with incident energy range of 300--450 MeV/u in inverse kinematics was performed at the R$^3$B/LAND setup at GSI Helmholtzzentrum f\"{u}r Schwerionenforschung in Darmstadt, Germany \cite{Pan16,Atar18,Fer18,Holl19}. While the measurements in Ref.~\cite{Pan16} are exclusive, those in Refs.~\cite{Atar18,Fer18,Holl19} provide the semi-inclusive cross sections and momentum distributions, in which various bound states of the residual nucleus are summed. The eikonal DWIA \cite{Pan16,Atar18,Holl19}, FAGS \cite{Fer18}, and TC \cite{Gom18} methods were applied to these experimental data and gave a generally similar conclusion that the reduction factors depend very weakly on the proton-neutron asymmetry, in contradiction with the much steeper asymmetry found in nucleon removal analysis. However, as was noted in Ref.~\cite{Gom18}, the $(p,pN)$ data analyzed by different reaction models exhibit some discrepancies due to choices of inputs and nonrelativistic treatments. These results also slightly underestimate the magnitude of $R_s$ given by \textit{ab initio} calculations \cite{Jen11,Cip15}. To clarify the inconsistencies between these models and give a reliable evaluation of the experimental data, it is of great interest to perform a careful DWIA analysis on the reduction factors with the GSI $(p,pN)$ data \cite{Pan16,Atar18,Fer18,Holl19}.           

The content of this paper is as follows. In Sec.~\ref{secformalism} the formulation of the standard partial-wave DWIA formalism is given. In Sec.~\ref{secresult} the cross sections and reduction factors are calculated for all published GSI data. The impact of several corrections in the DWIA framework on these observables are investigated. The transverse momentum distributions for the specific $^{12}$C($p$,$2p$)$^{11}$B case is also discussed. Finally, the summary is given in Sec.~\ref{secsum}.

\section{Formalism}
\label{secformalism}
The A($p$,$pN$)B knockout reaction is analyzed with the same partial-wave DWIA framework as in Ref.~\cite{Yos18}. Observables with superscript A are evaluated in the A-rest frame while those without the superscript are in the three-body center-of-mass (c.m.) frame, also called the G frame. The transition amplitude for the A($p$,$pN$)B reaction is given by
\begin{align}
T_{\bm{K}_0\bm{K}_1\bm{K}_2}^{nljm}
&=
\Braket{
\chi_{1,\bm{K}_1}^{(-)}\chi_{2,\bm{K}_2}^{(-)}
|t_{pN}|
\chi_{0,\bm{K}_0}^{(+)}\varphi^{nljm}
},
\label{eqtrans}
\end{align}
where $\chi_{i,\bm{K}_i} (i=0,1,2)$ are the distorted scattering wave functions of the $p$-A, $p$-B, and $N$-B systems, respectively. $\bm{K}_i$ is the momentum of particle $i$ in the G frame. As has been shown in Eqs.(3.8)-(3.15) of Ref.~\cite{Wak17}, when the kinematic coupling term in the exit channel Hamiltonian is approximated, the three-body scattering wave function can be separated into two two-body distorted-wave functions in Eq.~(1), where $\bm{K}_i$ can now be interpreted as the relative momentum. The superscripts $(+)$ and $(-)$ specify the outgoing and the incoming boundary conditions of these scattering waves, respectively. The relative single-particle wave function of the $N$-B system bound inside A is denoted as $\varphi^{nljm}$ where $n$, $l$, $j$, and $m$ are the principal quantum number, the orbital angular momentum, the total angular momentum, and its third component, respectively. $t_{pN}$ is a transition operator for the $p$-$N$ scattering, which is sometimes called an effective interaction \cite{FL85}. The absolute square of its matrix element is proportional to the $pN$ elastic-scattering cross section ${d\sigma_{pN}}/{d\Omega_{pN}}$.

Following the same theoretical treatment as in Ref.~\cite{Yos18}, upon disregarding the spin-orbit distortion, the momentum distribution (MD) is given by 
\begin{align}
\frac{d\sigma}{d\bm{K}_\mathrm{B}^\mathrm{A}}
=
&C_0 \int d\bm{K}_1^\mathrm{A}d\bm{K}_2^\mathrm{A}
\delta(E_f^\mathrm{A}-E_i^\mathrm{A})
\delta(\bm{K}_f^\mathrm{A}-\bm{K}_i^\mathrm{A}) \nonumber \\
&\times \frac{E_1 E_2 E_\mathrm{B}}{E_1^\mathrm{A} E_2^\mathrm{A} E_\mathrm{B}^\mathrm{A}} 
\frac{d\sigma_{pN}}{d\Omega_{pN}}
\sum_{m}(2\pi)^2
\lvert
\bar{T}_{\bm{K}_0\bm{K}_1\bm{K}_2}^{nljm}
\rvert ^2, \label{eq.md}
\end{align}
where
\begin{align}
C_0
&\equiv
\frac{E_0^\mathrm{A}}{(\hbar c)^2 K_0^\mathrm{A}}
\frac{f_{pN}}{(2l+1)}\frac{\hbar^4}{(2\pi)^3 \mu_{pN}^2}.
\end{align}
The factor $f_{pN}$ equals to 1 for $(p,pn)$ and $1/2$ for $(p,2p)$ reactions. $E_i$ and $E^{\rm A}_i$ are the total (relativistic) energies of particle $i$ in the G and A-rest frames, respectively. 

The G-frame $pN$ scattering cross section in Eq.~(\ref{eq.md}) is related to the one in the two-nucleon c.m. frame, which we term the t frame, through 
\begin{align}
\frac{d\sigma_{pN}}{d\Omega_{pN}}=
\eta^2\frac{d\sigma_{pN}^\textrm{\,t}}{d\Omega_{pN}^\textrm{\,t}},
\end{align}
where
\begin{align}
\eta=\left(\frac{E_1^\mathrm{t} E_2^\mathrm{t} E_0^\mathrm{t} E_N^\mathrm{t}}{E_1 E_2 E_0 E_N}\right)^{1/2}, \label{eq.mol}
\end{align}
is the M\o ller factor required for the transformation of $t_{pN}$ from the t frame to the G frame in relativistic kinematics \cite{Mol45,Ker59}. The total energy of the struck nucleon $E_N$ is determined by the momentum conservation of the two colliding nucleons as \cite{Yos18}
\begin{align}
E_N=\dfrac{\hbar^2}{2\mu_{NB}}\left[\bm{K}_1+\bm{K}_2-\dfrac{(A+1)}{A}\bm{K}_0\right]^2.
\end{align}

The reduced transition amplitude in Eq.~(\ref{eq.md}) is given by
\begin{align}
\bar{T}_{\bm{K}_0\bm{K}_1\bm{K}_2}^{nljm}
&=
\int d\bm{R}\,\chi_{1,\bm{K}_1}^{*(-)}(\bm{R})\,
\chi_{2,\bm{K}_2}^{*(-)}(\bm{R})\,
\chi_{0,\bm{K}_0}^{(+)}(\bm{R})\,
\nonumber \\
&\times
\varphi^{nljm}(\bm{R}) e^{-i\bm{K}_0\cdot\bm{R}/A}.
\end{align}
For the bound-state wave function $\varphi^{nljm}$ generated by a local potential, the effect of nonlocality is taken into accounted in the interior region by multiplying $\varphi^{nljm}$ by the Perey factor \cite{Per63}
\begin{align}
F_\textrm{PR}(R)=C_\textrm{PR}\left[1-\dfrac{\mu_{N\mathrm{B}}}{2\hbar^2}\beta^2 V_{N\mathrm{B}}(R)\right]^{-1/2},
\end{align}
where the nonlocality range $\beta=0.85$ fm for nucleon \cite{Per62}, and $V_{N\mathrm{B}}$ is the single-particle binding potential. The factor $C_\textrm{PR}$ is chosen so that the modified bound-state wave function is normalized to unity. Similarly, for a scattering wave function obtained from a Dirac phenomenology optical potential (OP), the relativistic velocity-dependent term modifies the wave function by what we refer to as the Darwin factor \cite{Arn81,Ham90} 
\begin{align}
F_\textrm{DW}(R)=\left[\dfrac{E_i+U_S(R)-U_V(R)}{E_i}\right]^{1/2},
\end{align}
where $U_S$ and $U_V$ are the scalar and vector potentials in the Dirac equation, respectively. This Darwin factor is regarded as a kind of nonlocality correction and has been well known to be very important in order to fully take into account relativistic effect in $(e,e'p)$ reactions \cite{Udi95}. 

The cylindrical transverse momentum distributions (TMD) are obtained from the MD as
\begin{align}
\frac{d\sigma}{dK_{\mathrm{B}b}^\mathrm{A}}
&=
2\pi\int dK_{\mathrm{B}z}^\mathrm{A} K_{\mathrm{B}b}^\mathrm{A}
\frac{d\sigma}{d\bm{K}_\mathrm{B}^\mathrm{A}}.
\end{align}
The integrated single-particle cross section is then calculated from the TMD as
\begin{align}
\sigma_\textrm{sp}
&=
\int \frac{d\sigma}{dK_{\mathrm{B}b}^\mathrm{A}}dK_{\mathrm{B}b}^\mathrm{A}. \label{eq.sp}
\end{align}

\section{Results and discussion \label{secresult}}
In this section, we applied the DWIA framework described in the preceding section to the GSI $(p,pN)$ data \cite{Pan16,Atar18,Fer18,Holl19} to evaluate the single-particle cross section and deduce the reduction factor. The impact of nonlocality corrections, the M\o ller factor, and energy dependence of final-state OP on the $R_s$ are clarified.   

\subsection{Numerical inputs}
\label{subsecinput}

We perform DWIA calculations for all 18 published $(p,pN)$ cases of the R$^3$B/LAND setup \cite{Pan16,Atar18,Fer18,Holl19}. In addition, we also perform calculations with several other choices of inputs to estimate the theoretical uncertainty on reduction factors. All calculations in this work adopt the relativistic treatment of the kinematics, which is essential to reproduce the correct MD \cite{Yos18}. The nonlocality correction is taken into account in our calculation through the use of the Perey factor for the single-particle bound state and Darwin factor for distorted waves.

For the distorting potential of the $p$-A, $p$-B, and $N$-B systems, we use the EDAD2 parameter set of the Dirac phenomenology \cite{Coo93}. Calculations using the EDAD1, EDAD3 \cite{Coo93}, and the ``democratic'' EDAD \cite{Coo09} Dirac OP sets give a difference of 10\% at most, which is consistent with the one observed in normal kinematics \cite{Wak17}. 

The single-particle wave function of the struck nucleon is obtained from a Woods-Saxon potential with central and spin-orbit components defined in the same manner as in Refs.~\cite{Gade08,Gom18}. For both components, a diffuseness $a=0.7$~fm is used for all the cases. The radius parameter is adjusted following the prescription $\langle r^2 \rangle=[A/(A-1)]\langle r^2 \rangle_\textrm{HF}$, where $\langle r^2 \rangle_\textrm{HF}$ is the single-particle mean square radius of the Hartree-Fock calculation with the Skyrme SkX interaction \cite{Bro98}. The depth $V_\textrm{so}=6$ MeV is fixed for the spin-orbit term while the central one is adjusted to reproduce the experimental separation energies. This choice of binding potential gives a difference in the $(p,pN)$ cross section within 10\% compared with the one used in the $(e,e'p)$ analysis \cite{Kra01,Wak17}. Based on this result and the investigations of Refs.~\cite{Gade08,Gom18} on the ambiguity of different effective nucleon-nucleon (\textit{NN}) interactions used in the Hartree-Fock calculation, we adopt an uncertainty of 10\% for single-particle wave functions.    

For the $pN$ elementary cross section, we employ the one generated by the $t$-matrix parametrization of Franey and Love \cite{FL85} with a final-energy prescription, which has been suggested to be the best approximation for the half-off-shell amplitude \cite{Red70}. Different choices of the on-shell approximation such as initial-energy and average-energy prescriptions give an uncertainty of 2\% for $(p,2p)$ and 8\% for $(p,pn)$ processes. The large discrepancies in neutron knockout processes are due to the asymmetric shape of the $pn$ scattering angular distribution that makes it more sensitive to the choice of energy prescription. Note that other choices of the \textit{NN} cross section such as those from the Reid93 potential \cite{Reid93} in TC \cite{Gom18}, the CD-Bonn potential \cite{Mac01} in FAGS \cite{Fer18}, or from experimental database fitting \cite{Arn03} in the eikonal DWIA \cite{Aum13} give essentially the same \textit{NN} cross sections up to 350 MeV.  

The theoretical spectroscopic factor for each state of the bound residual core is the same as in Refs.~\cite{Gom18,Holl19}, which is computed by the standard SM with the WBT interaction \cite{War92} and includes the c.m. correction \cite{Die74}. Since many of the considered nuclei are weakly bound, the use of SF calculated from the SM, compared with the IPM limit, provides a more proper description of single-particle-strength fragmentation near the Fermi level. 

\subsection{Reduction factors} \label{sec.redfac}

\begin{table*}[htbp]
	\caption{Experimental cross sections $\sigma_\textrm{exp}$ \cite{Atar18,Pan16,Fer18,Holl19}, calculated ones $\sigma_\textrm{th}$, and reduction factors $R_s$. See the text for details.}
	\begin{ruledtabular}
		\begin{tabular}{lcddc}
			Reaction & \multicolumn{1}{l}{$E_\textrm{beam}$ (MeV/u)} & \multicolumn{1}{c}{$\sigma_\textrm{th}$ (mb)} & \multicolumn{1}{c}{$\sigma_\textrm{exp}$ (mb)} & \multicolumn{1}{c}{$R_s$} \\ \hline
			$^{10}$C($p$,$pn$)$^{9}$C  & 386   & 12.95 & 16.3(22)[14]      & 1.26(29)   \\
			$^{11}$C($p$,$2p$)$^{10}$B & 325   & 15.68 & 18.2(9)[10]       & 1.16(19)   \\
			$^{11}$C($p$,$pn$)$^{10}$C & 325   & 14.07 & 17.0(15)[21]      & 1.21(27)   \\
			$^{12}$C($p$,$2p$)$^{11}$B & 398   & 22.04 & 19.2(18)[12]      & 0.87(16)   \\
			$^{12}$C($p$,$pn$)$^{11}$C & 398   & 27.43 & 30.0(32)[27]      & 1.09(23)   \\
			$^{13}$O($p$,$2p$)$^{12}$N & 401   & 5.77  & 5.78(91)[37]      & 1.00(22)   \\
			$^{14}$O($p$,$2p$)$^{13}$N & 351   & 13.28 & 10.23(80)[65]     & 0.77(13)   \\
			$^{15}$O($p$,$2p$)$^{14}$N & 310   & 18.07 & 18.92(182)[120]   & 1.05(19)   \\
			$^{16}$O($p$,$2p$)$^{15}$N & 451   & 27.78 & 26.84(90)[170]    & 0.97(15)   \\
			$^{17}$O($p$,$2p$)$^{16}$N & 406   & 9.16  & 7.90(26)[50]      & 0.86(14)   \\
			$^{18}$O($p$,$2p$)$^{17}$N & 368   & 20.01 & 17.80(104)[113]   & 0.89(15)   \\
			$^{21}$O($p$,$2p$)$^{20}$N & 449   & 5.58  & 5.31(23)[34]      & 0.95(15)   \\
			$^{21}$N($p$,$2p$)$^{20}$C & 417   & 3.25  & 2.27(34)          & 0.70(14)   \\
			$^{21}$N($p$,$pn$)$^{20}$N & 417   & 38.87 & 48.52(404)        & 1.25(23)   \\
			$^{22}$O($p$,$2p$)$^{21}$N & 414   & 6.90  & 6.01(41)          & 0.87(14)   \\
			$^{22}$O($p$,$pn$)$^{21}$O & 414   & 36.24 & 39.24(234)        & 1.08(19)   \\
			$^{23}$O($p$,$2p$)$^{22}$N & 445   & 4.97  & 4.93(96)          & 0.99(24)   \\
			$^{23}$O($p$,$pn$)$^{22}$O & 445   & 50.05 & 54.0(108)         & 1.08(28)   \\
		\end{tabular}%
		\label{tab.result}%
	\end{ruledtabular}
\end{table*}%

The results of our DWIA calculations are presented in Table ~\ref{tab.result}. Because of the semi-inclusive nature of the data concerned, the calculation results are the sum of the cross sections corresponding to several bound-state configurations of the residual nucleus. The beam energy in the middle of the target is shown in the second column. The third column indicates the theoretical cross sections $\sigma_\textrm{th}=\sum C^2S \times \sigma_\textrm{sp}$, where $\sigma_\textrm{sp}$ is the single-particle cross section of a specific configuration. The fourth column shows the experimental cross section, with statistical (round brackets) and systematic (square brackets) uncertainties, corresponding to the beam of $^{10-12}$C \cite{Pan16,Holl19}, $^{13-21}$O \cite{Atar18} and $^{21}$N,$^{22,23}$O \cite{Fer18}. The reduction factor $R_s=\sigma_\textrm{exp}/\sigma_\textrm{th}$ is given in the last column.  

\begin{figure}[htbp]
	\centering
	\includegraphics[width=0.48\textwidth]{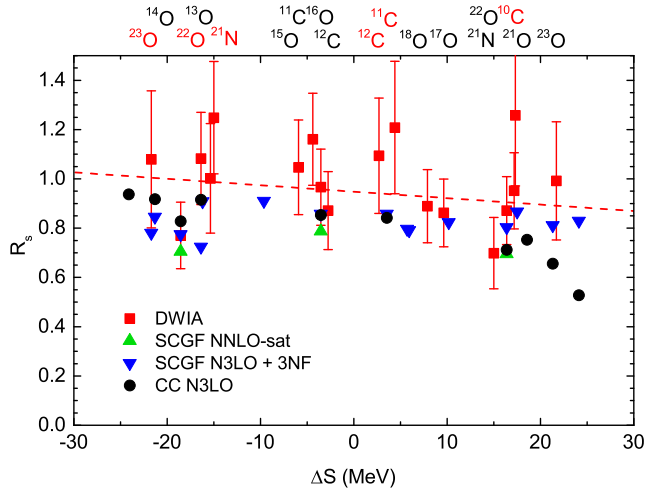}
	\caption{
		Reduction factors deduced from $(p,pN)$ reactions using DWIA compared with the reduced SFs calculated from CC \cite{Jen11} (black circles) and SCGF \cite{Cip15,Atar18} with different interactions (blue and green triangles). The black (red) texts in the header label the isotope whose proton (neutron) is knocked out. See the text for details. 
	}
	\label{fig.main-theory}
\end{figure}

The reduction factors as a function of the proton-neutron asymmetry $\Delta S$ is shown in Fig.~\ref{fig.main-theory}. The value calculated from the present DWIA analysis is indicated by red squares with error bars propagated from the experimental uncertainties reported in Refs.~\cite{Pan16,Atar18,Fer18,Holl19} and the theoretical uncertainties estimated in Sec.~\ref{secformalism}, which is about 14\%--16\%. The total relative uncertainties for the extracted reduction factor are ranging from 15\% to 25\%, which is similar to those reported in the recent systematic $(p,d)$ analysis \cite{Xu19}. In general, a trend of the reduction factor about 0.9--1.0 with a very weak asymmetry dependence is observed. Properly taking the uncertainties into account, the reduction factors $R_s=0.87(16)$ for $^{12}$C and $R_s=0.97(15)$ for $^{16}$O are overestimated compared with the $(e,e'p)$ \cite{Lap93} results of $R_s=0.57(6)$ and $R_s=0.65(5)$, respectively. By performing a linear function fitting, the $\Delta S$ dependence of $R_s$ is obtained as $R_s=0.947(36)-2.6(27)\times 10^{-3}\Delta S$ with a reduced $\chi^2/N$ of 0.74. As discussed in Sec.~\ref{sec.reaction}, the close-to-unity reduction factor does not necessarily mean that the quenching effect observed in $(p,pN)$ reactions is weak but rather indicates a fundamental problem in current reaction models.   

The reduction factors from DWIA are compared with the reduced SFs, which is the ratio of the SF to the IPM limit, from \textit{ab initio} self-consistent Green’s function (SCGF) \cite{Atar18,Cip15} and coupled-cluster (CC) \cite{Jen11} models. We note that all of these values are presented as a function of \textit{experimental} $\Delta S$. The CC calculation \cite{Jen11} uses the chiral \textit{NN} interaction at next-to-next-to-next-to-leading order (N3LO) \cite{Ent03} with the cutoff at $\Lambda_\textit{NN}=500$ MeV while the SCGF of Ref.~\cite{Cip15} uses the same \textit{NN} interaction in addition to an NNLO three-nucleon force \cite{Roth12} with $\Lambda_{3N}=400$ MeV. We compare also with the result of the more recent SCGF calculation in Ref.~\cite{Atar18} based on the NNLO-sat \cite{Eks15} for the \textit{NN} interaction, which is more optimized for the mass region in that study. The present DWIA calculation shows a reasonable agreement with \textit{ab initio} results, especially about the weak dependence of the trend on proton-neutron asymmetry.

\subsection{Comparison with $\bm{(e,e'p)}$ results and other reaction models} \label{sec.reaction}
\begin{figure}[htbp]
	\centering
	\includegraphics[width=0.48\textwidth]{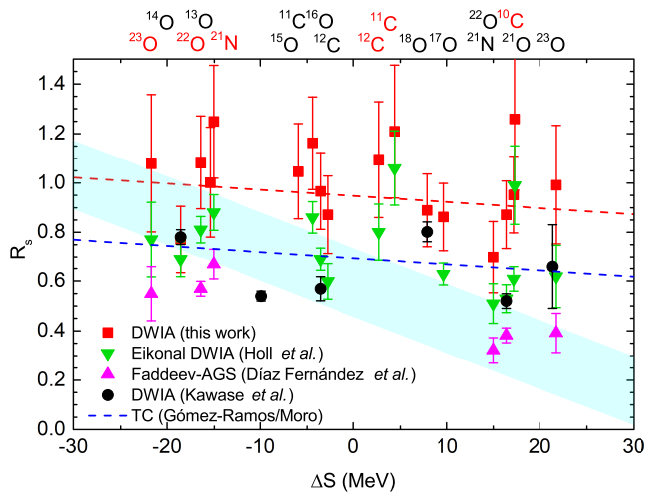}
	\caption{
		Same as Fig.~\ref{fig.main-theory} but compared with other $(p,pN)$ analysis performed with the eikonal DWIA \cite{Atar18} (green triangles), FAGS \cite{Fer18} (pink triangles), TC \cite{Gom18} (blue dashed line) for the GSI data, and similar partial-wave DWIA for the RIKEN/RCNP data \cite{Kaw18} (black circles). The results from Refs.~\cite{Gom18,Kaw18} are those performed with Dirac OP. The blue shaded band indicates the trend observed in the analysis of nucleon removal reaction with composite targets \cite{Gade08,Tos14}.
	}
	\label{fig.main-reaction}
\end{figure}

Figure \ref{fig.main-reaction} shows the reduction factors from the present DWIA calculation in comparison with those analyzed by other theoretical reaction models. The slope of the reduction factor obtained by the present analysis is in excellent agreement with the value deduced by TC model with fixed-energy EDAD2 OP \cite{Gom18}. It is also consistent (within the uncertainty range) with the slopes observed in Refs.~\cite{Atar18,Fla18}.

The values of the reduction factor given by DWIA overestimate those reported by $(e,e'p)$ analysis \cite{Lap93} and other reaction models such as TC \cite{Gom18} and the eikonal DWIA \cite{Atar18}. This means the present DWIA calculation gives a smaller cross section compared with these. However, the better agreement between the $(e,e'p)$ results and those given by other $(p,pN)$ analyses of the same GSI data will be, in fact, caused by the lack of several crucial corrections in the reaction models used in these analyses. We will return to this point below. 

The overshooting of the $(e,e'p)$ results means that the DWIA cross section is smaller than the observed value. This indicates that some contributions from higher-order processes such as multistep scattering or channel coupling are included in the GSI data, especially in the large recoil momentum region. In fact, a smaller reduction factor around 0.7 is observed from the analysis of $(p,2p)$ data measured at RIKEN/RCNP \cite{Kaw18} using the same DWIA framework as in this study. An important feature of the measurement at RIKEN/RCNP is a very
constrained kinematics corresponding to the quasifree condition. This supports the conclusion that the lack of higher-order effects will be the main reason for the underestimation of the GSI data with the current DWIA calculation.

The discrepancies between the current DWIA results and those using other reaction models can be explained by several factors. First, the nonlocality corrections in the single-particle and scattering wave functions are not presented in the TC and eikonal DWIA models. Second, the TC calculation \cite{Gom18} uses the energy-independent optical potentials evaluated at $E_\textrm{beam}/2$. On the other hand, the large discrepancy with the FAGS method \cite{Fer18} is due to two reasons pointed out in Ref.~\cite{Gom18}, the lack of relativistic kinematics in the FAGS framework as shown in \cite{Yos18} and different choices of optical potentials used in Ref.~\cite{Fer18}. The latter has been confirmed by recent FAGS calculation with a more proper OP for the energy range considered, where the cross section has been reduced by almost 30\% \cite{Cre19}.    

\begin{figure}[htbp]
	\centering
	\includegraphics[width=0.48\textwidth]{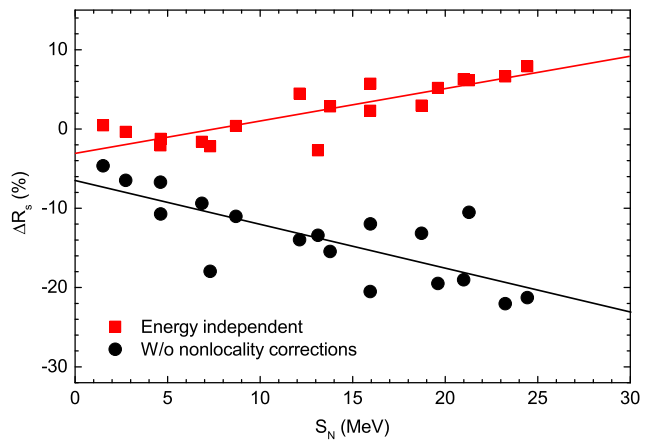}
	\caption{
		The relative difference with respect to the reference DWIA results as a function of knocked-out-nucleon separation energy. The red squares and black circles represent the DWIA calculations with energy-dependent potentials and without nonlocality corrections, respectively.  
	}
	\label{fig.deltaNL}
\end{figure}

We further investigate the possible source of discrepancies between different reaction models used in $(p,pN)$ studies. Impacts of the lack of nonlocality corrections and energy-dependent potentials on the reduction factors are illustrated in Fig.~\ref{fig.deltaNL}. The relative difference $\Delta R$ is evaluated with respect to the ``reference'' DWIA result shown in Sec.~\ref{sec.redfac}. As seen from the lines in Fig.~\ref{fig.deltaNL}, these effects are strongly associated with the separation energy and affect the reduction factor though the cross section in the opposite ways. For the weakly bound nucleus, the single-particle wave function is more extended and since the nonlocality correction only affects the interior of the wave function, its effect gets weaker with decreasing separation energy. On the other hand, since the energy-independent potentials of the outgoing nucleons are evaluated at half the incident beam, the deviation from energy-dependent OP is minimized for the small separation energy, where the kinematics most resembles the quasifree condition. Some deviations from both lines are because of the cross section for each case is the sum of single-particle wave functions with different orbitals. In general, by neglecting both the nonlocality correction and the energy dependence of the OP, DWIA calculations are expected to have a similar magnitude to those from TC \cite{Yos18}.

\begin{figure}[htbp]
	\centering
	\includegraphics[width=0.48\textwidth]{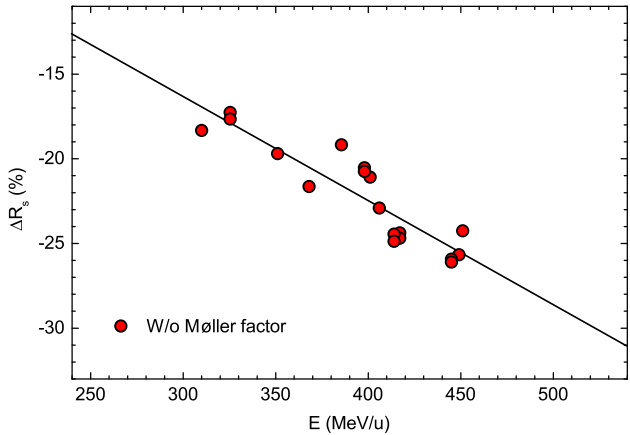}
	\caption{
		The relative difference between the reference DWIA results and those without the M\o ller factor as a function of beam energy.
	}
	\label{fig.deltaMol}
\end{figure}

Finally, the effect of the M\o ller factor is presented in Fig.~\ref{fig.deltaMol}. This factor, shown in Eq.~(\ref{eq.mol}), has a relativistic origin \cite{Mol45,Ker59,Udi95} and therefore is directly related to the incident energy. For the considered energy range of 300--450 MeV, the M\o ller factor can contribute about 18\%--26\% to the reduction factor, which also explains the small magnitude of the nonrelativistic FAGS model. We note that, although the TC model does not explicitly include the M\o ller factor, its consistency with the DWIA model as found in \cite{Yos18} suggests that the M\o ller-factor effect has been implicitly included in the relativistic treatment of TC \cite{Mor15}. 

Although the lack of nonlocality correction, energy dependence in OP, and the M\o ller factor may cancel with the lack of higher-order effects in some models and give a better agreement with $(e,e'p)$ results, they also obscure the true nature of the problem. As we have seen, the effects of these corrections are highly dependent on the separation energy of the single-particle wave function and beam energy, so more proper inclusion of these corrections will be very essential for future knockout studies.   

\subsection{Transverse momentum distribution}

In inverse kinematics nucleon knockout reactions, a comparison between the measured and calculated momentum distribution of the residual nucleus can reveal a lot of information about the reaction mechanism as well as the validity of the theoretical model. We consider the cylindrical TMD of $^{12}$C($p$,$2p$)$^{11}$B \cite{Pan16}, which is some of the highest resolution data from the R$^3$B/LAND experiments.

\begin{figure}[htbp]
	\centering
	\includegraphics[width=0.46\textwidth]{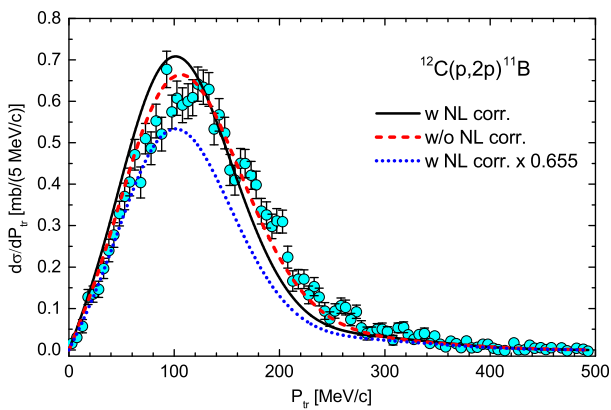}
	\caption{
		Cylindrical transverse momentum distribution of the $^{12}$C($p$,$2p$)$^{11}$B reaction. The experimental data are taken from Ref.~\cite{Pan16}. The DWIA calculations with and without the nonlocality corrections are presented as the black solid and red dashed lines, respectively. The blue dotted line represents the result with nonlocality corrections scaled down by a factor of 0.655.
	}
	\label{fig.12C}
\end{figure} 

The possible absence of higher-order effects is pronounced in the comparison between the DWIA result (black solid line) and the experimental data \cite{Pan16} shown in Fig.~\ref{fig.12C}. The calculation results are the sum of the theoretical TMD corresponding to the bound states of the $^{11}$B core multiplied by the reduction factor $R_s$ of 0.87. One sees that the calculation result including the nonlocality corrections underestimates the experimental data in the momentum region about 150--300 MeV/$c$. We note that this undershooting of theoretical prediction is also observed in the FAGS calculation for the same data using OP fit to elastic-scattering data at the proper energy range \cite{Cre19}. Moreover, a similar discrepancy with the data in the high-momentum region found in the Cartesian TMD calculated with TC \cite{Gom18} implies that some contributions are also not properly accounted for in other reaction models. Because of the success of the DWIA framework for the same system in the quasifree condition \cite{Wak17}, the discrepancies reinforce our claim that they originate from the higher-order effects that take place when the recoil momentum becomes high. 

The red dashed line represents the DWIA result without nonlocality correction, i.e., neglecting the Perey and Darwin factors. The corresponding $R_s$ is 0.77. Although this line may seem to improve the result, we emphasize that such a prescription is inappropriate and inconsistent with the known properties of $(p,pN)$ scattering as reviewed in
Ref.~\cite{Wak17}. The blue dotted line is the DWIA result with nonlocality corrections multiplied by 0.66, which is the ratio of the $R_\textrm{s}$ in Table \ref{tab.result} to the one determined by $(e,e’p)$ \cite{Lap93}. If the $(e,e’p)$ $R_s$ is correct, the blue dotted line should agree with the experimental data. Therefore, it is clear that some additional contributions are necessary to reproduce the data.   

\section{Summary}
\label{secsum}
We have performed an analysis on all the published data to date for the $(p,pN)$ reaction in inverse kinematics by the R$^3$B collaboration by using the standard partial-wave DWIA formalism. Our study focuses on evaluating the source of ambiguity in DWIA calculations of the reduction factor and investigating the discrepancies between various reaction models currently used for inverse kinematics $(p,pN)$ data.
 
Our study suggests a very weak dependence of the reduction factor on the proton-neutron asymmetry $\Delta S$. This result is consistent with previous analyses on $(p,pN)$ using TC, eikonal and partial-wave DWIA \cite{Gom18,Atar18,Kaw18,Holl19} and \textit{ab initio} calculations \cite{Cip15,Atar18,Jen11}. 

More importantly, the present study suggests that the lack of a proper treatment of higher-order effects may considerably affect the cross section at kinematics far from the quasifree condition. That effect, which does not manifest in experiments performed around the recoilless condition, becomes more crucial in the semi-inclusive type of integrated cross-section measurements like those carried out at the GSI R$^3$B/LAND setup. However, it is mostly hindered by the lack of essential corrections such as nonlocality, relativistic, M\o ller factor, and energy-dependent OP as well as the considerably large uncertainty from various choices of distorting potentials and single-particle wave functions. The proper inclusion of these corrections in the future $(p,pN)$-reactions analyses will be required. Furthermore, unless measurements with restricted kinematics similar to that in Ref.~\cite{Kaw18} are performed, a reaction model that takes into account higher-order processes will be necessary. 

Recently, a consistent description of the nonlocality in bound and scattering wave functions for $(e,e'p)$ has been done with the nonlocal dispersive optical model approach \cite{Atk18}. The incorporation of such treatment in the DWIA framework for proton-induced nucleon knockout reactions is in progress and will be reported elsewhere.            

\begin{acknowledgments}
We thank Mario G\'{o}mez-Ramos and Matthias Holl for providing essential details of their calculations. We also thank Tetsuo Noro for valuable discussions on the DWIA calculation. One of the authors (N.T.T.P) would like to thank the Research Center for Nuclear Physics, Osaka University for their hospitality during his stay in which this work was initiated. He also wants to thank the Vietnam MOST for its support through the Physics Development Program Grant No. {\DJ}T{\DJ}LCN.25/18. The computation was carried out with the computer facilities at the Research Center for Nuclear Physics, Osaka University. This work was supported in part by Grants-in-Aid of the Japan Society for the Promotion of Science (Grants No. JP16K05352), and the RCNP Young Foreign Scientist Promotion Program.
\end{acknowledgments}


\end{document}